\documentclass[prl,twocolumn,showpacs,preprintnumbers,superscriptaddress,amsmath,amssymb]{revtex4}

\usepackage{epsf,color,graphicx}
\usepackage[dvips]{epsfig}
\usepackage{graphicx}

\begin{document}

\title{Exciton polaritons in two-dimensional photonic crystals}

\author{D.~Bajoni}\email[]{daniele.bajoni@unipv.it}
\affiliation{CNISM and Dipartimento di Elettronica, Universit\`{a} degli Studi di Pavia, via Ferrata 1, 27100 Pavia, Italy}
\author{D.~Gerace}
\author{M.~Galli}
\affiliation{CNISM and Dipartimento di Fisica ``A. Volta,'' 
Universit\`{a} degli Studi di Pavia, via Bassi 6, 27100 Pavia, Italy}
\author{J.~Bloch}
\author{R.~Braive}
\author{I.~Sagnes}
\author{A.~Miard}
\author{A.~Lema\^{i}tre}
\affiliation{CNRS-Laboratoire de Photonique et de Nanostructures,
Route de Nozay, 91460 Marcoussis, France}
%\author{Marco Malvezzi}
%\affiliation{CNISM UDR Pavia and Dipartimento di Elettronica, Universit\`{a} degli studi di Pavia, via Ferrata 1, 27100 Pavia, Italy}
\author{M.~Patrini$^2$}
\author{L.~C.~Andreani$^2$}

\date{\today}

\begin{abstract}
Experimental evidence of strong coupling between excitons confined in a quantum well and the
photonic modes of a two-dimensional dielectric lattice is reported. Both resonant scattering and
photoluminescence spectra at low temperature show the anticrossing of the polariton branches,
fingerprint of strong coupling regime. The experiments are successfully interpreted in terms of
a quantum theory of exciton-photon coupling in the investigated structure. These results show that the polariton dispersion can be tailored by properly varying the photonic crystal lattice parameter, which opens the possibility to obtain the generation of entangled photon pairs through polariton stimulated scattering.

\end{abstract}

\pacs{71.36.+c, 42.70.Qs, 78.55.Cr, 42.30.Kq, 71.35.-y}
%\keywords{}

\maketitle

% INTRODUCTION
The strong coupling regime between light and matter is characterized by a reversible and coherent 
exchange of energy between a single material oscillator and a single mode of the electromagnetic 
field. A particular case is  when excitons confined in a semiconductor Quantum Well (QW) are spectrally and spatially resonant with the mode of a vertical semiconductor microcavity, e.g. in structures similar to the Vertical Cavity Surface Emitting Laser \cite{VCSEL}. If the coherent light-matter coupling overcomes excitonic and photonic dissipation rates, the strong coupling regime can be achieved in these structures \cite{Weisbuch,Skolnick}. As a result, exciton-photon hybrid quasiparticles named \emph{microcavity polaritons} are formed. 

Microcavity polaritons have bosonic statistics, fast interaction with the electromagnetic field, and strong optical nonlinearities related to their excitonic parts. These properties have been exploited to obtain a wealth of nonclassical phenomena in the solid state: coherent and macroscopically occupied matter-wave states \cite{Kasprzak,Balili}, optical spin Hall effect \cite{Leyder}, and superfluidity \cite{Amo} among others. Their strong optical nonlinearities have been used to demonstrate low-power parametric oscillations of matter-waves \cite{Savvidis,Stevenson,Tignon,Saba}: considerable interest has been devoted to polariton parametric scattering with the goal of realizing a semiconductor-based, monolithic and micron-sized source of entangled photon pairs. However, quantum correlation experiments are usually hindered by the great intensity difference between signal and idler beams, which is intrinsic to the dispersion of polariton branches in planar microcavities \cite{Karr}: possible solutions have been sought by modifying the microcavity geometry \cite{Tignon,Pillars}. 

%Here we report the first experimental evidence of strong coupling between excitons confined in a semiconductor QW and the photonic modes of a two-dimensional photonic crystal. 
Photonic crystals can be used to tune the photonic mode dispersion by 
suitably modifying the sample design \cite{Joannopoulos}. Photonic crystals in the strong coupling regime give the unique possibility to engineer the dispersion of polariton branches. New ways to achieve phase matching for parametric scattering, e.g. to obtain signal and idler beams of comparable intensity, can thus be envisioned, opening the possibility of measuring quantum correlations between signal and idler polaritons.
Early experimental evidence of polaritons in photonic crystals has
been reported by using polymers sputtered on gratings
\cite{Ishihara,shimada02}. 
Polymers undergoing strong coupling have very strong oscillator strengths
but do not show optical nonlinearities. Although theoretically
proposed \cite{Gerace}, polaritons in photonic crystals have never
been reported in semiconductors, where nonlinearities are very
high \cite{Saba}. The main problem is that, in many semiconductor
systems, and in particular in GaAs-based samples, patterning the QW
results in a severe reduction of the exciton lifetime by
nonradiative recombination at hole sidewalls \cite{Vuckovic}, and
consequently in the loss of strong coupling \cite{Koch}.

Here we report the first experimental evidence of strong coupling between excitons confined in a semiconductor QW and the photonic modes of a two-dimensional photonic crystal. To avoid issues related to nonradiative
recombination, we adopt an original design: the photonic
lattice is spatially separated from the QWs so that polaritons
only experience the periodic potential through their photonic part, while
leaving the QW intact. 

\begin{figure}[t]
\begin{center}
\includegraphics[width=0.9\columnwidth]{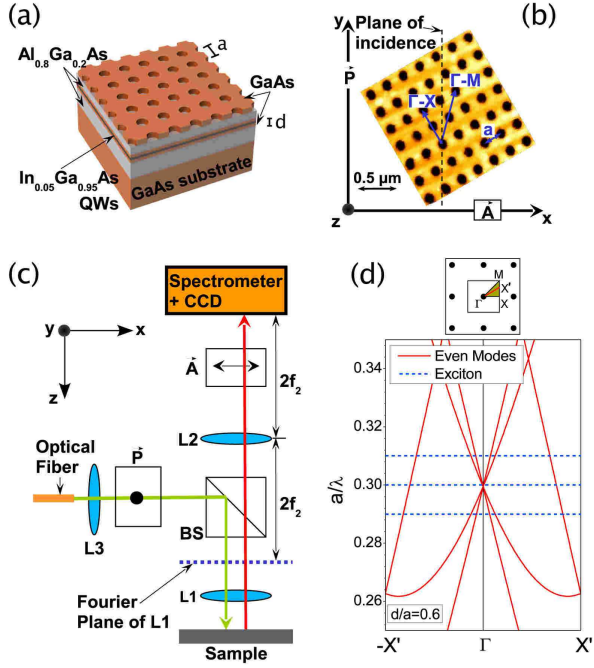}
\caption{(color online) (a) Schematic sample structure. Relevant parameters
are the waveguide core thickness, $d$, and the lattice constant,
$a$. (b) Atomic Force Microscope
image of the surface of a sample with $a$=250 nm. The main lattice
directions, as well as the plane of incidence and the orientation
of the polarizers are shown. (c) Experimental set-up: 
the green line is the incoming beam (laser or white light) and the red 
line is the output beam (PL or reflected light), while $\vec{\text{A}}$ and
$\vec{\text{P}}$ are two linear polarizers, L$_1$,
L$_2$ and L$_3$ are lenses and f$_2$ is the focal length of L$_2$,
BS is a beam splitter. 
(d) Schematic Brillouin zone and guided mode dispersion in the experimental plane of 
incidence, notice that the energy scale is in adimensional units. The modes are calculated for a symmetric planar GaAs waveguide with 
between AlGaAs claddings, with dielectric constants 
$\varepsilon_{\mathrm{core}}=12.97$ and $\varepsilon_{\mathrm{clad}}=9.5$, respectively.
The exciton resonance for three different lattice parameters is represented 
with dashed lines.}\label{Fig1}
\end{center}
\end{figure}

% SAMPLE AND EXP SET-UP
The sample, schematically shown in Fig.~\ref{Fig1}a, was grown by molecular beam 
epitaxy on a GaAs substrate. A 140 nm
thick Al$_{0.8}$Ga$_{0.2}$As cladding was first deposited, followed by a 148 nm thick GaAs core with three 8 nm thick In$_{0.05}$Ga$_{0.95}$As QWs %(separated by 12 nm spacers) 
at its center, a second 140 nm thick
Al$_{0.8}$Ga$_{0.2}$As cladding and finally a 100 nm GaAs top layer.
The top layer was patterned by inductively coupled plasma etching \cite{Sagnes}
with a square lattice of circular air holes (see Fig.~\ref{Fig1}b):
areas with different lattice parameters $a$=245, 250, 255, and 260
nm were defined. The nominal etch depth is 120 nm. The periodic corrugation yields a dispersion-folding on the {\it guided} modes of the 
slab waveguide within the first Brillouin zone (Fig.~\ref{Fig1}d, upper panel), making them 
{\it radiative} around normal incidence ($\Gamma$-point). 
Such folding depends on the lattice parameter, $a$: for different lattice parameters, 
the resonance condition between the QW exciton and the photonic modes changes 
within the dispersion diagram. This is the essence of the polariton dispersion 
engineering discussed in the following, and it is shown in the calculated mode 
dispersion in Fig.~\ref{Fig1}d.

The sample was kept at low temperature in a He-cooled cold finger
cryostat. The experimental set-up is outlined in Fig.~\ref{Fig1}c. 
To probe the elementary excitations of the system two
different techniques were used: photoluminescence (PL) and
resonant scattering (RS) of white light. This latter approach
relies on sending a linearly polarized white light beam along the $y$-axis (by
the polarizer $\vec{\text{P}}$ in Fig.~\ref{Fig1}c) and
analyzing the reflected light in crossed polarization through a
second polarizer along $x$ ($\vec{\text{A}}$). 
The photonic modes of the sample have
a symmetry that can be even (i.e. having the
electric field mainly polarized in the QW plane, TE-like) or odd
(i.e. with the electric field mainly polarized perpendicular to
the QW plane, TM-like). A crucial point in the present experiment
is that the plane of incidence is tilted by $\phi=30^{\circ}$ with
respect to the $\Gamma$X lattice direction (Figs.~\ref{Fig1}b,c)
and is not a mirror plane of the structure.  While most of
the reflected light keeps the polarization specified by the
polarizer $\vec{\text{P}}$, when the incoming beam is resonant
with a photonic/polaritonic mode a small amount of light is
coupled to the opposite polarization, i.e. parallel to 
$\vec{\text{A}}$. Thus, the scattering signal is resonantly
enhanced \cite{Williams} and the modes appear as peaks over an
almost vanishing background.  The RS signal is a
measure of the extinction of light due to its coupling with the sample 
and can be used to probe the dispersion of the modes in a wide spectral range: 
in this sense, it is analogous to
the resonant Brillouin scattering technique for measuring the
dispersion of exciton-polaritons in bulk semiconductors
\cite{Ulbrich}. 
% For both PL and RS
% measurements, the collected signal is then dispersed by a grating
% spectrophotometer with a LN$_2$ cooled CCD.
PL spectra are obtained on the same set-up employing a laser pump at 1550 meV. 
Both the laser pump and the white light beam are focused on a 20
$\mu$m-diameter spot on the sample surface by a microscope objective (L$_1$
in Fig.~\ref{Fig1}a), that is also used to collect the
emitted/reflected light. The plane of incidence is determined by
the entrance slit of the spectrophotometer (parallel to $y$ in the frame of Fig.~\ref{Fig1}) 
and resolution in the angle of incidence ($\vartheta$) is obtained by directly 
imaging the Fourier plane of L$_1$ onto the slit.

\begin{figure}[t]
\begin{center}
\includegraphics[width= 0.9\columnwidth]{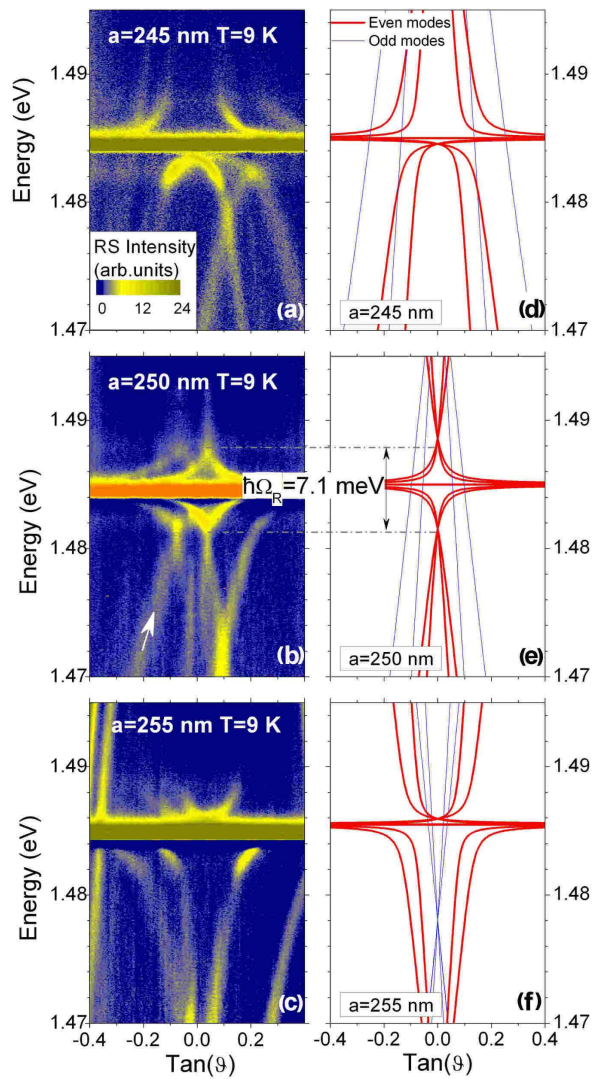}%[width= 0.42\textwidth]{Fig2.eps}
\caption{(color online) 
Experimental RS spectra for samples with lattice
constants (a) $a=245$, (b) 250, and (c) 255 nm, respectively. 
(d), (e), (f)  Corresponding calculated dispersion curves for even 
(red) and odd (blue) modes.} \label{Fig2}
\end{center}
\end{figure}

\begin{figure}[t]
\begin{center}
\includegraphics[width=0.9\columnwidth]{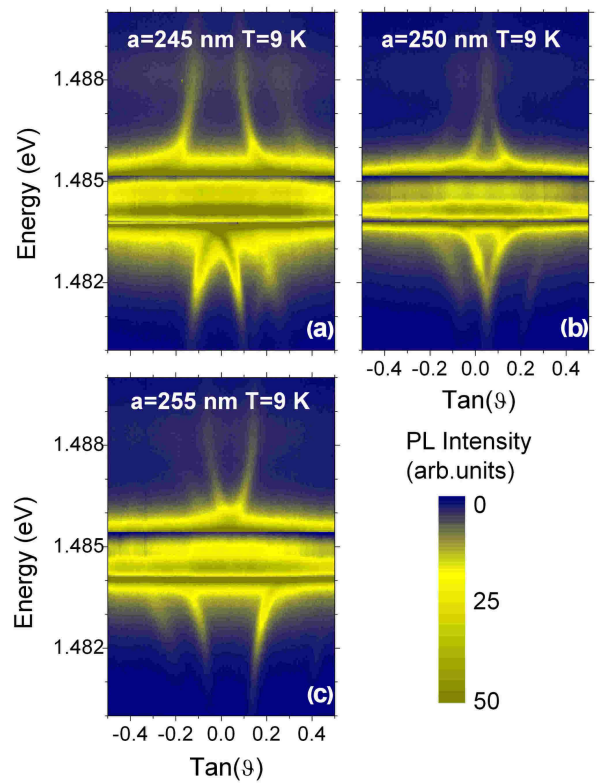}%[width=0.4\textwidth]{Fig3.eps}
\caption{(color online)  Experimental PL spectra for samples with lattice
constants (a) $a=245$, (b) 250, and (c) 255 nm, respectively.  The
intensity has been divided by a factor of 5 in a window of $\sim 1.3$ meV around the
exciton resonance.} \label{Fig3}
\end{center}
\end{figure}

Polariton states, i.e. mixed radiation-matter quasiparticles, are expected to occur 
when the dimensionality of photon states is equal to (or, possibly, smaller
than) that of the exciton \cite{Andreani-Varenna}. Given the energies $E_X$ and $E_{Ph}$ of the uncoupled exciton and photonic modes and the respective dissipation rates $\gamma_X$ and $\gamma_{Ph}$, 
the energies of the polariton eigenstates are simply described by \cite{Skolnick}:
\begin{equation} 
\begin{small}
\begin{aligned}
E_{{LP}/{UP}} &= \frac{1}{2}\left( E_X + i \gamma_X + E_{Ph} + i \gamma_{Ph} \right)\\ & \mp \frac{1}{2} \sqrt{ \left( \hbar\Omega_R \right)^2 + \left( E_X + i \gamma_X - E_{Ph} - i \gamma_{Ph} \right)^2 },  
\end{aligned}
\label{eqantix}
\end{small}
\end{equation}
where $\hbar\Omega_R /2$ is the coupling constant between excitons and photons (half of 
the vacuum Rabi splitting). Equation \ref{eqantix} implies that polaritons anticross in reciprocal space, where the uncoupled modes would cross instead: the observation of such anticrossing is a clear signature of the strong coupling regime. 

% RS RESULTS
Low temperature RS
measurements as a function of the incidence angle, $\vartheta$, 
are shown in Figs.~\ref{Fig2}a,b,c  for samples with $a$=245, 250, and 255 nm, respectively.
All spectra display an anticrossing between the dispersive modes
and the exciton line at 1.485 eV, proving the
occurrence of a strong coupling regime. Both the upper and lower polariton branches
are clearly visible on either side of the exciton and the
polariton linewidth remains well below 1 meV in all spectra; the
measured Rabi splitting is $\simeq 7$ meV, comparable to what
has been reported for planar microcavities. Notice that, without coupling with the
exciton, the modal dispersion would be linear, as discussed below. Figure 2 highlights that the original shape of polariton dispersions in such photonic crystals : for instance the diamond-like shape of the modes in Fig. 2b is completely different from the S-shaped dispersion of microcavity polaritons \cite{Skolnick}. Moreover the polariton dispersion dramatically dependends on the lattice constant: changing $a$ by only
2\% (5 nm) substantially reshapes the branches. 
%This result opens the way towards the unique possibility of engineering
%the polariton dispersion by varying the sample design. 
%Notice also that the dispersion measured for $a=255$ nm,
%with a minimum of the upper branch at $\vartheta$=0 and the lower
%branch tending to the exciton at high $\vartheta$, is reversed for
%$a$=245 nm, which has a maximum for the lower branch at
%$\vartheta$=0. 

% THEORY AND DISCUSSION
We model the system under investigation by calculating the
photonic modes for the planar waveguide in the two-dimensional 
lattice as in Fig.~\ref{Fig1}d. 
% the dispersion of bare even (TE-like)
% and odd (TM-like) guided modes, when folded along the main
% symmetry lines in reciprocal space, is shown in Fig.~2d.
Exciton-photon coupling is described by a full quantum
formulation, as  detailed in Ref.~\cite{Gerace}: 
the Rabi splitting in the polariton mode dispersion
depends on the overlap between 
the guided modes of the slab waveguide and the QW exciton 
envelope function, as well as on the oscillator strength per unit area~\cite{nota}. 
The resulting polariton dispersions are
reported in Figs.~\ref{Fig2}d,e,f, for the same lattice
parameters as in Figs.~\ref{Fig2}a,b,c. 
Overall, the details of the polariton modes are
complicated by the tilted plane of incidence ($30^{\circ}$) with
respect to the $\Gamma$X direction, which removes photonic mode
degeneracies.
Two photonic branches for each parity can be distinguished at the
$\Gamma$-point ($\vartheta=0^{\circ}$), but only even modes
strongly couple to the excitons. 
There is a good agreement between theory and experiment: the calculated Rabi
splitting for these structures is $\hbar\Omega_R \simeq 7.1$ meV and compares very
well with the experimentally determined value of $\sim 7$ meV ,
as it can be easily seen by directly comparing Figs.~\ref{Fig2}b and e.
Notice the presence of additional polariton branches in the measurements, which are 
not reproduced by the theory (an instance is highlighted by a white arrow in Fig.~\ref{Fig2}b). 
These additional modes appear only when the plane of incidence does not correspond to 
a high symmetry direction of the lattice ($\Gamma$M or $\Gamma$X) and may be related 
to finite lateral size of the samples: further theoretical and experimental investigation is 
underway to confirm this interpretation.

\begin{figure}[t]
\begin{center}
\includegraphics[width=0.9\columnwidth]{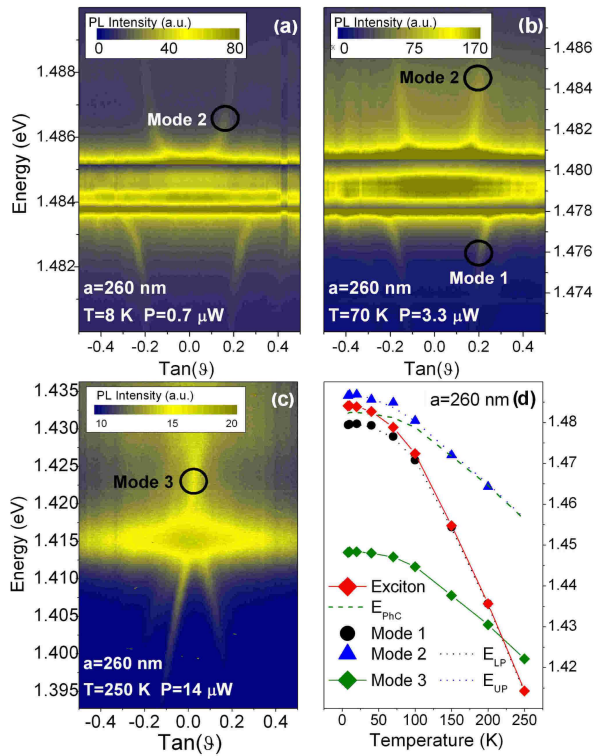}%[width=0.42\textwidth]{Fig4.eps}
\caption{(color online) Experimental PL spectra for a sample with
a=260 nm at temperatures (a) T=8, (b) 70 and (c) 250 K, respectively. 
In panels (a) and (b),
the intensity in a window of $\sim1.3$~meV around the exciton
resonance has been divided by a factor of 5. 
(d) Energies of the
exciton (red tilted square), mode 1 (black circles, black circle in panel b), 
mode 2 (black triangles, black circle in panel b) and mode 3 (green tilted square, 
green circle in panel c) as a function of temperature. The green
dashed line is the uncoupled photonic mode at
$\vartheta$=10$^{\circ}$ (i.e. $\text{Tan} \vartheta\sim 0.18$).} \label{Fig4}
\end{center}
\end{figure}

% PL RESULTS
Measured PL spectra are shown in Figs. \ref{Fig3}a,b and c. 
At such low temperatures, only polariton
states lying a few meV from the exciton are enough populated to
efficiently contribute to the PL signal, and the energy scale is consequently 
expanded in the figure.
Polariton dispersion lines in emission are exactly equivalent to those
reported in RS, confirming the occurrence of strong coupling in these samples. 
The exciton line at 1.485 eV is much
more intense than the polariton lines. This is mainly due to two 
reasons: first, in the present experiment we collect the PL signal on a
much wider region than the excitation spot, so that we measure
also the signal coming from diffused excitons outside
the patterned region. Second, only exciton states
with symmetry close to the photonic mode are strongly coupled, while the majority of states remain
in weak coupling with the electromagnetic field \cite{Gerace}. 
% In Fig.~\ref{Fig2} a) one of the modes (highlighted by an arrow)
% is very weakly coupled to the exciton, as the dispersion crosses
% the exciton line. It is attributed to a mode with odd symmetry
% (TM-like), which cannot couple to the in-plane polarized QW
% exciton; as a further evidence, odd modes are completely invisible
% in PL experiments.

%In Fig.~\ref{Fig2} a) one of the modes with odd symmetry is
%visible and is highlighted by a black arrow: these modes are very
%weakly coupled to the exciton (which is in-plane polarized) and
%indeed the mode dispersion crosses the exciton line; as further
%evidence odd modes are completely invisible in PL experiments.

% The bright modes visible at high $\vartheta$ in Fig.~\ref{Fig2}c)
% are Wood anomalies, \textit{i.e.} light scattered at the interface
% between the cladding and the patterned layer (see also
% Fig.~\ref{Fig3}d): these modes are unrelated to the exciton and
% invisible in the PL.

% ANALYSIS
To evidence the difference between strong and weak coupling in our samples, PL measurements with increasing temperature ($T$) are
plotted in Figs.~\ref{Fig4}a,b c for a sample with $a$=260 nm. At T=70 K the sample is still in strong coupling, and shows the same
anticrossing as at T=8 K 
(Figs.~\ref{Fig4}a,b). Above 80 K the Rabi splitting is progressively reduced and at T=250 K
(Fig.~\ref{Fig4}c) strong coupling is lost and the photonic modes cross the exciton resonance. As an illustration, fig. 4d shows the energy variation with temperature of the exciton and of  three modes chosen around the
exciton at $\vartheta=10^{\circ}$ (modes 1 and 2, see black circles in Figs. 4a,b) and at low energy at
$\vartheta=0^{\circ}$ (mode 3 see black circle in Fig. 4c) extracted from the spectra.
Since the exciton redshift is stronger than that of the photonic modes, increasing temperature
changes the detuning between exciton and modes. The exciton and mode 3 
cross above 200 K, which evidences they are in weak coupling. On the other hand, modes
1 and 2 anticross at 45 K, and are thus in strong coupling. Their energies can be
fitted by using Eq. 1, in which the uncoupled photonic mode, $E_{Ph}$, is assumed to have the
same dependence on the temperature as mode 3. This loss of strong coupling at high temperature can be understood considering the exciton dephasing
rate 
$\gamma_X$ in Eq. \ref{eqantix}: increasing temperature means increasing $\gamma_X$ 
until the term under square root becomes negative, i.e. excitons dephase in a time 
$~1/ \gamma_X$ before a single Rabi oscillation can be completed. 
%The exciton energy 
%and the energies of three modes chosen around the exciton at $\vartheta=10^{\circ}$ 
%(modes 1 and 2, see black circles in Figs.~\ref{Fig4}a,b) and at low energy 
%at $\vartheta=0^{\circ}$ (mode 3 see black circle in Fig.~\ref{Fig4}c) extracted from the 
%spectra are shown in Fig.~\ref{Fig4}d. 
%Since the exciton redshift is stronger than that of the photonic modes, increasing 
%temperature changes the detuning between exciton and modes. 
%The exciton and mode 3 clearly cross above 200 K, which evidences they are in weak 
%coupling. On the other hand, modes 1 and 2 clearly anticross at 45 K, and are thus in strong 
%coupling. Their energies can be fitted by using Eq.~\ref{eqantix}, in which 
%the uncoupled photonic mode, $E_{Ph}$, is assumed to have the same dependence on the 
%temperature as mode 3.

% CONCLUSIONS
In conclusion, we have shown  the strong coupling regime of
quantum-well excitons in two-dimensional photonic crystals
through both resonant scattering and photoluminescence experiments
at low temperature. 
The present results will open new directions for polariton research:
the possibility of engineering phase matching is a considerable step towards 
achieving a compact and integrable
solid-state source of entangled photon pairs.
Further applications can be envisioned for polaritons in large band gap materials.
Although the strong coupling regime has been already reported at room
temperature in GaN \cite{Semond,Grandjean}, ZnO \cite{Morkoc} and
CuCl \cite{Nakayama}, fabrication of high quality microcavities
for these materials is often impractical due to the large number
of required layers. Using planar photonic crystals, and in
particular the present design that leaves the active region
intact, will turn to be effective for obtaining strong coupling with high quality
optical modes at room temperature.

% ACKNOWLEDGEMENTS
The authors acknowledge M. Malvezzi for fruitful discussions and F. Manni for participation in 
the early part of the work. This work was supported by CNISM
through the "Innesco" initiative. Financial support from Fondazione Cariplo 
through project 2007-5259 is acknowledged.

\end{document}